\documentclass[titlepage,12pt]{article}
\usepackage{amssymb,psfig,epsfig}
\newcommand {\real} {{\rm Re}}
\newcommand {\imag} {{\rm Im}}
\newcommand{\hst}{\hspace*{0.3cm}}

\begin{document}
\begin{titlepage}
\noindent 
PITHA 03/08 \hfill \\

\vspace{2cm}
\begin{center}
{\LARGE {\bf Semileptonic decays of polarised top quarks: $V + A$
admixture and QCD corrections}} \\
\vspace{2cm}
{\bf Werner Bernreuther\footnote{{\it E-mail address:} 
\tt{breuther@physik.rwth-aachen.de}},
Michael  F\"ucker\footnote{supported by  D.F.G. SFB/TR9}
and Yoshiaki Umeda\footnote{supported by BMBF contract 05 HT1PAA/4}}
\par\vspace{1cm}
Institut f.\ Theoretische Physik, RWTH Aachen, 52056 Aachen, Germany\\
\par\vspace{3cm}
{{\bf Abstract:}}\\
\parbox[t]{\textwidth}
{The semileptonic decays of polarised
top quarks are analysed for a general chirality-conserving
$tbW$ vertex. We
calculate  double differential
distributions for the charged lepton  and the neutrino
to order $\alpha_s$ in the QCD coupling.
We present these QCD corrections 
in terms of compact
parameterisations that should be useful for the future investigation
of the structure of the top decay vertex on the basis of large
data samples.}
\end{center}
\vspace{2cm}
PACS number(s): 12.60.Cn, 13.88.+e, 14.65.Ha
\end{titlepage}
%
\setcounter{footnote}{0}
\renewcommand{\thefootnote}{\arabic{footnote}}
\setcounter{page}{1}
\noindent
So far the experimental information about the decays of
top quarks is not very detailed. Data from the Tevatron are 
consistent with the expectation that these decays are governed by the
$V-A$ charged current interactions of the standard model (SM); but
because of the size of the 
present experimental errors  sizable  new physics effects
can not be excluded (c.f. for instance \cite{CDFhel}).

In view of the   extremely large mass of the top quarks
and the circumstance that
they do  not hadronise these particles are excellent probes
of new interactions that may uncover  at energies of a few hundred GeV.
There is a vast literature on the phenomenology of such interactions
in top quark decays (for overviews, see for instance 
\cite{Beneke:2000hk,Aguilar-Saavedra:2001rg}). One possibility is a
small  $V+A$ admixture to the standard left-handed $t \to b$
current as is predicted for instance by 
$SU(2)_L \times SU(2)_R \times U(1)$ extensions of the standard model.
The measured branching ratio of the radiative weak decay 
$b \to s \gamma$ provides a stringent constraint 
\cite{Cho:zb,Fujikawa:1993zu,Hosch:1996wu,Larios:1999au} 
on the coupling strength $\kappa_R$ of such a right-handed  admixture:
$|{\kappa}_R|\lesssim 0.04.$ Of course, one may envisage
contrived scenarios where other contributions to $b \to s \gamma$
lead to cancellations that invalidate this bound -- in any case,
a direct search for such a coupling in $t$ and $\bar t$ decays is
indispensable. Sensitive  tests require polarised
top quarks. At the LHC the single top production processes
will yield large samples of polarised $t$ and $\bar t$ quarks, and
it has been estimated \cite{Espriu:2001vj,delAguila:2002nf}  
that a statistical sensitivity 
$\delta{\kappa}_R \simeq 0.06$ can be reached.  From
$t \bar t$ pair production at a high-luminosity
linear $e^+e^-$ collider a sensitivity $\delta{\kappa}_R \simeq 0.03$
will be  feasible \cite{Schm,Boos:1999ca,Aguilar-Saavedra:2001rg}.

Energy and angular distributions
for polarised top quark decays with a $V+A$ admixture in the
$tbW$ vertex were investigated in a number of articles, including
\cite{Ma:ry,Jezabek:1994zv,Nelson:1997vs,Grzadkowski:2000nx}. In ref.
\cite{Jezabek:1994zv} semileptonic $t$ decays were analysed and it
was pointed out that the polarisation-dependent part of the neutrino
distribution (which can be determined experimentally by measuring the
missing momentum) is very sensitive to $\kappa_R$. The order
$\alpha_s$ QCD corrections for the $V-A$ part of the
charged current \cite{Czarnecki:1990pe,Czarnecki:1994pu}
were incorporated in \cite{Jezabek:1994zv}.
In view of the above-mentioned expected sensitivity to $\kappa_R$ at 
future colliders one should note that taking into account QCD
corrections in future data analysis is mandatory, because they can mimic a
small $V+A$ admixture.

In this paper we extend the work of \cite{Jezabek:1994zv} 
in that we compute the order $\alpha_s$ QCD corrections to both the
$V-A$ and $V+A$ Born contributions to $t\to b \ell \nu_{\ell}$. 
Moreover we take the finite width of the intermediate $W$ boson and
the non-zero mass of the $b$ quark into account. The latter is
important in the  case of a small $V+A$ admixture, because terms
linear in $\kappa_R$ in the lepton distributions, which we compute
below, require a chirality flip of the $b$ quark.
For small $\kappa_R$,
say  $|{\kappa}_R| < 0.1$, the terms proportional
to $\kappa_R m_b/m_t$ are of
the same order of magnitude as the contributions proportional to
$\kappa_R^2$. Furthermore we compute a T-odd triple correlation
which  is generated by  a non-standard CP-violating phase that may 
be present in the general chirality-conserving $tbW$ vertex.

We consider the semileptonic decay of 
spin-polarised top quarks to order $\alpha_s$ in the QCD coupling,
which amounts to studying the reactions
\begin{eqnarray}
t & \rightarrow & b (p_b) + \ell^+(p_\ell) + {\nu}_{\ell}(p_\nu) \, ,
\nonumber \\
t & \rightarrow &  b (p_b) + \ell^+(p_\ell) + {\nu}_{\ell}(p_\nu) +g(p_g)
\, .
\label{reac}
\end{eqnarray}
All momenta refer to the rest frame of the top quark. 
As far as the $tbW$ vertex is concerned we use the generalised interaction
\begin{equation}
{\cal L}_{tb} = -\frac{g}{2\sqrt{2}}V_{tb}\,
\bar{t}\gamma^{\mu}(\alpha-\beta\gamma_5)
b \, W_\mu \;\; + \;\; {\rm h.c.} \, ,
\label{Ltb}
\end{equation}
where $\alpha, \beta$ are complex couplings.  A right-handed 
 admixture to 
the $V-A$ charged current interaction of the
SM ($\alpha = \beta =1$) will be parameterised by choosing 
$\alpha=1+\kappa_R,$ $\beta=1-\kappa_R.$ 
In the calculations below
we neglect the lepton masses but take
the mass of the $b$ quark and the finite width of the intermediate $W$
boson into account. 

In the calculation of the differential decay distributions for
(\ref{reac}) we have  performed  the quark
wave function renormalisations that remove the ultraviolet
divergencies in the on-shell scheme.  The infrared divergencies are
canceled  using a  standard phase space slicing procedure. 
The phase space of the four-particle final state in the reaction
$t\to b \ell \nu_{\ell} g$ is split into two disjoint regions where
the  (scaled) energy $x_g=2E_g/m_t$ of the
gluon is smaller and larger than an arbitrary, but
small separation parameter $x_{min}$, respectively: $1=\Theta(x_{min} - x_g) + 
\Theta(x_g - x_{min}).$  Integrating the respective squared matrix
element over the phase space of the soft gluon $(x_g\leq  x_{min})$ and
adding  the result to the order $\alpha_s$ squared matrix element for
the three particle final state $ b \ell \nu_{\ell}$ yields the infrared
finite differential decay distribution $d\Gamma_{B+V+soft}$ (B = Born,
V = virtual) and $d\Gamma_{hard},$  which
 describes  radiation  of ``resolved'' gluons $(x_g >  x_{min}).$

For checks of possible deviations from the $V-A$ law in the $tbW$
vertex useful observables are the double differential energy-angle
distributions $d\Gamma/dx_id\cos\theta_i$ $(i=\ell, \nu)$ for the
charged lepton and for the neutrino, where  $x_i=2E_i/m_t$ and
$\theta_i$ is the angle between the three-momentum of the lepton $i$ and
the unit
vector $\hat{\bf s}$ that specifies the polarisation 
direction of the ensemble of top
quarks in the $t$ rest frame. The degree of polarisation
is denoted by $S=|{\bf s}|.$ Adding the contributions from
 $d\Gamma_{B+V+soft}$ and from $d\Gamma_{hard}$, these
distributions can be put into the following form:
\begin{equation}
\frac{d \Gamma}{dx_\ell d\cos\theta_{\ell} } 
= |V_{tb}|^2 \frac{g^4m_t}{8}
    \left [( F^{\ell n}_{0} + C_F\frac{\alpha_s}{\pi} 
 F^{\ell n}_{1}) +(F^{\ell s}_{0} + C_F\frac{\alpha_s}{\pi} 
F^{\ell s}_{1} )
S\,\cos\theta_{\ell} \right ] \, ,
\label{xellth}
\end{equation}
\begin{equation}
\frac{d \Gamma}{dx_\nu d\cos\theta_{\nu}} 
= |V_{tb}|^2 \frac{g^4m_t}{8}
\left [ (F^{\nu n}_{0} +  C_F\frac{\alpha_s}{\pi} F^{\nu n}_{1})
+ (F^{\nu s}_{0} +  C_F\frac{\alpha_s}{\pi}F^{\nu s}_{1})
S\, \cos\theta_{\nu}  \right ] \, ,
\label{xnuth}
\end{equation}
where $C_F=4/3$ and 
$0 \leq x_{\ell}, x_{\nu}\leq (m_t^2-m_b^2)/m_t^2$. 
The corresponding distributions
for the anti-top quark are obtained by changing the
sign of the term proportional to $\cos\theta_i$.
From these distributions the corresponding 
$2\times 2$ top decay spin density
matrices  can be extracted  in straightforward 
fashion\footnote{These
density matrices enter the calculations of the (differential) cross sections
that describe
$t$ and/or ${\bar t}$ production and decay in the on-shell approximation.}.

The $F^{ij}_a$ 
 ($i=\ell,\nu,$ $j=n,s,$ and $a=0,1$) which are functions of
 $x_{\ell}$ and  $x_{\nu}$, respectively,  can be decomposed according
to the contributions from the vector and axial vector
couplings  $\alpha,\beta:$
\begin{equation}
F^{ij}_{a} = |\alpha|^2 F^{ij}_{a,\alpha}
            +\real \, ({\alpha^*\beta}) 
              F^{ij}_{a,\alpha\beta}
           +|\beta|^2 F^{ij}_{a,\beta} \, .
\label{eq:aa-ab-bb}
\end{equation}
The following relations hold between the functions
that appear  in the charged lepton distribution and those of the
corresponding neutrino distribution:
\begin{eqnarray}
F^{\ell n}_{a,\alpha}=F^{\nu n}_{a,\alpha} \, , \qquad
F^{\ell n}_{a,\beta}=F^{\nu n}_{a,\beta} \, ,\qquad
F^{\ell n}_{a,\alpha\beta}= - \, F^{\nu n}_{a,\alpha \beta} \, , 
\nonumber \\
F^{\ell s}_{a,\alpha}= - \, F^{\nu s}_{a,\alpha} \, , \qquad
F^{\ell s}_{a,\beta}= - \, F^{\nu s}_{a,\beta} \, ,\qquad
F^{\ell s}_{a,\alpha\beta}=  F^{\nu s}_{a,\alpha \beta} \, , 
\label{relf01}
\end{eqnarray}
where $a=0,1.$ 
For the sake of presenting  compact expressions we
 write down the lowest order functions of
the charged lepton distribution in terms of one-dimensional integrals
($r=\alpha,\beta, \alpha\beta$):
\begin{equation}
F^{\ell j}_{0,r} = \frac{1}{512\pi^3}
\int_{x_{b,min}}^{x_{b,max}}\!\!\! dx_b \, 
D_W(x_b) {\tilde F}^{\ell j}_{0,r}(x_{\ell},x_b) \, ,
\label{intF}
\end{equation}
where 
 $D_W=  ((1-x_b-{\hat{m}_W}^2+{\hat{m}_b}^2)^2+
{\hat{m}_W}^2{\hat{\Gamma}_W}^2)^{-1},$ 
${\hat{m}_W}= m_W/m_t,$ ${\hat{\Gamma}_W}=\Gamma_W/m_t,$
${\hat{m}_b}= m_b/m_t,$
$ x_{b,min}=(1-x_{\ell} + {\hat{m}_b}^2/(1-x_{\ell})),$
$x_{b,max}= (1+{\hat{m}_b}^2),$  and 
\begin{eqnarray}
{\tilde F}^{\ell n}_{0,\alpha}&=& -2 {\hat{m}_b}^3 + 
{\hat{m}_b}^2 (x_b-2) + 
        2 {\hat{m}_b} (x_b-1) - x_b^2 + 
         x_b(3 - 2x_\ell) - 
        2 (x_\ell-1)^2  \, , \nonumber \\
{\tilde F}^{ln}_{0,\alpha\beta}&=& 2(1+{\hat{m}_b}^2-x_b)
(2-x_b-2x_\ell) \, , \nonumber \\
{\tilde F}^{ln}_{0,\beta}&=& 2 {\hat{m}_b}^3 + {\hat{m}_b}^2 (x_b-2) - 
        2 {\hat{m}_b} (x_b-1) - x_b^2 + 
         x_b(3 - 2x_\ell) -  2 (x_\ell-1)^2  \, ,\nonumber \\
{\tilde F}^{ls}_{0,\alpha}&=& -(1+{\hat{m}_b}^2-x_b)(2
  {\hat{m}_b}^2+x_b(x_\ell-2)+2(x_\ell-1)^2+2{\hat{m}_b} x_\ell) \, ,
\nonumber \\
{\tilde F}^{ls}_{0,\alpha\beta}&=& 4(1+{\hat{m}_b}^2-x_b)^2+
2(1-{\hat{m}_b}^2-x_b)(x_b-4)x_\ell-4(x_b-2)x_\ell^2-4x_\ell^3 \, , 
\nonumber\\
{\tilde F}^{ls}_{0,\beta}&=&-(1+{\hat{m}_b}^2-x_b)(2
  {\hat{m}_b}^2+x_b(x_\ell-2)+2(x_\ell-1)^2-2 {\hat{m}_b} x_\ell) \, . 
\end{eqnarray}
The corresponding functions for the neutrino distribution are obtained
with the relations (\ref{relf01}). Using $\alpha, \beta = 1\pm
\kappa_R$ it is easy to see that those  terms in the Born
distributions
which are linear in $\kappa_R$ are accompanied by a factor
${\hat{m}_b}$ (or ${\hat m_b}^3$).  

The functions $ F^{ij}_{1,\alpha},  F^{ij}_{1,\alpha\beta},$
and $F^{ij}_{1,\beta}$ which are induced by the order $\alpha_s$ corrections
are shown in Figs. 1 and 2.  
We use $m_t=174.3$ GeV, $m_b=4.8$ GeV, $m_W=80.42 $ GeV, and $\Gamma_W=2.12$
GeV. For SM couplings $\alpha=\beta = 1$ and for the narrow width
approximation ${\hat{\Gamma}_W} \to 0$ 
we can numerically compare our results with 
those of \cite{Czarnecki:1994pu} and we find agreement.

The size of the QCD corrections is typically 6 to 8 percent  for $x_i <  0.9$
and somewhat larger at the upper end of the spectrum. For $x_i \to 1$
the curves show 
the developoment of the well-known logarithmic singulariy 
\cite{Czarnecki:1990pe} which
can be removed by exponentiation.
Rather than presenting the exact analytical
formulae for the functions $ F^{ij}_{1,\alpha},  F^{ij}_{1,\alpha\beta},$
and $F^{ij}_{1,\beta}$, which are quite lengthy, we give them in
terms of simple parameterisations determined by fits. 
We required the fits to be better than 4$\%$ in the range 
$0 \leq x_{\ell}, x_{\nu}\leq 0.98$. This
requirement can be satisfied by using  a rational function for 
$x_{\ell}, x_{\nu}\leq 0.218$ 
and a polynomial
of at most 8th order for $x_{\ell}, x_{\nu}> 0.218$. 
This number is the location of the peaks
in Figs. 1 and 2  which correspond to the maximum of $D_W$ and the shape
of the functions of $F_1$ around this peak arise as follows.
The contributions to $F_1$
from  $d\Gamma_{V+soft}$ and $d\Gamma_{hard}$ have opposite signs.
The virtual and soft corrections start growing steeply at a slightly
lower value of  $x_i$ than the term from ``hard''  gluon radiation 
and the former
level off slightly earlier, at  $x_i\simeq 0.218$, where the increase
of  $d\Gamma_{hard}$ is still rather steep.
We have checked numerically that this does
not depend on the choice of the (small) separation parameter $x_{min}.$

\begin{table}[tp]
\begin{center}
\begin{tabular}{|c||l|l|l|l|l|l|}\hline
&\hst$F^{\ell n}_{1,\alpha}$\hst &\hst$F^{\ell n}_{1,\beta}$\hst 
&\hst$F^{\ell n}_{1,\alpha\beta}$\hst 
&\hst$F^{\ell s}_{1,\alpha}$\hst &\hst$F^{\ell s}_{1,\beta}$\hst 
&\hst$F^{\ell s}_{1,\alpha\beta}$\hst 
\\
\hline\hline
$A_0$$[10^{-7}]$& $-$8.06 & $-$10.54 & $-$14.37
                & $-$7.55 & $-$9.86  & $-$15.25 \\ \hline
$A_1$$[10^{-5}]$&\hst5.50 &\hst5.50 &\hst4.96
                &\hst5.24 &\hst5.24 &\hst5.18 \\ \hline
$B_0$  & $-$0.2853   & $-$0.3087    & $-$1.6414
       & $-$1.1058   & $-$1.2267    & $-$1.6760 \\ \hline
$B_1$  &\hst3.3781   &\hst3.6545    &\hst22.6920
       &\hst17.7909  &\hst19.7949   &\hst23.4286  \\ \hline
$B_2$  & $-$16.3481  & $-$17.902   & $-$131.1934  
       & $-$122.4657 & $-$136.8286 & $-$136.6096  \\ \hline
$B_3$  &\hst40.0362  &\hst44.3232  &\hst408.7910
       &\hst467.4425 &\hst523.4506 &\hst428.1320 \\ \hline
$B_4$  & $-$53.1628  & $-$59.4165  & $-$741.6436
       & $-$1083.4018& $-$1213.9973& $-$781.1829 \\ \hline
$B_5$  &\hst36.4636  &\hst41.0886  &\hst784.5948
       &\hst1563.7047&\hst1751.0343&\hst831.2825 \\ \hline
$B_6$  & $-$10.1118  & $-$11.4767   & $-$448.9931
       & $-$1374.8811& $-$1536.9986& $-$478.5832 \\ \hline
$B_7$  & & &\hst107.4585
       &\hst674.4979 &\hst752.1898 &\hst115.2665  \\ \hline
$B_8$  &  &  & 
       & $-$141.6121 & $-$157.4561 & \\ \hline
\end{tabular}
\end{center}
\caption{The coefficients which determine the fits (\ref{eq:fit})
to the order $\alpha_s$ QCD contributions to the charged lepton
distribution (\ref{xellth}).  $A_0$ and $A_1$ are
given in units of $10^{-7}$ and $10^{-5},$ respectively.}
\label{table:fit}
\end{table}

In the fits we  use the generic form
\begin{equation}
F_1 =
\Theta (0.218-x) \frac{A_0}
{(x-0.218)^2+A_1} + \Theta (x-0.218) \sum_{k=0}^{8} B_k x^k \, ,
\label{eq:fit}
\end{equation}
where $\Theta (x)$ denotes the step function.
The coefficients $A_k, B_k$ that specify the respective functions
are given in Table 1.  The corresponding functions for the neutrino
distribution are obtained with the help of (\ref{relf01}).

The polarisation dependent part of the neutrino distribution,
$F^{\nu s}\equiv  F^{\nu s}_{0} +  C_F\frac{\alpha_s}{\pi}F^{\nu s}_{1},$ 
is the term in the above distributions 
that is most sensitive to a coupling 
$\kappa_R$ \cite{Jezabek:1994zv}, especially for neutrino
energies $x_{\nu} \lesssim 0.6.$ For instance, for $\kappa_R = 0.1$ 
the function $F^{\nu s}$ deviates from its SM value  
by about $4 \%$ for  $0.25 \lesssim x_{\nu} \lesssim 0.6,$ 
and for $x_{\nu} \lesssim 0.25$ the deviations are even larger.
In order to obtain maximal
sensitivity to $\kappa_R$ in future data analyses one may use
likelihood functions that can be derived \cite{Schm}
from (\ref{xellth}) and (\ref{xnuth}).

Finally we consider the possibility that ${\imag}(\alpha^*\beta) \neq
0.$ This happens in a natural way in $SU(2)_L \times SU(2)_R \times
U(1)$
extensions of the SM, where the Higgs sector is sufficiently complicated to
allow for a CP-violating phase in the $L-R$ gauge boson mixing matrix
already at tree level. After diagonalisation of this mass
matrix this phase is transported
to the charged current interaction (\ref{Ltb}).
An observable to check for such a phase is the expectation
value of the T-odd triple product  
${\cal O} = \hat{\bf p}_{\ell}\cdot (\hat{\bf p}_b \times
\hat{\bf s})$,
where $\hat{\bf{p}}_{b,\ell}$ are the directions of flight of the
b quark and the lepton, respectively, in the top quark rest frame.
(At this order in the QCD coupling the decay amplitude
has  no absorptive part  that could generate $<{\cal O}> \neq 0.$)
We have to order $\alpha_s$:
\begin{equation}
< {\cal O} > 
= \frac {1}{\Gamma_{SL}} (\int {\cal O} \, d\Gamma_{B+V+soft}
\;  +\; \int {\cal O} \, d\Gamma_{hard} \, ) \, ,
\label{eq:eps}
\end{equation}
where $\Gamma_{SL}$ denotes the  order $\alpha_s$ partial decay width for
(\ref{reac}). Because the deviations of the moduli of the parameters
$\alpha$ and $\beta$ from their SM values are not expected to be large
we use ${\Gamma^{SM}_{SL}}$ for the normalisation in
(\ref{eq:eps}). We use ${\Gamma^{SM}_{SL}}$ = 0.168 GeV and 0.153 GeV
to  lowest order and to order
$\alpha_s,$ respectively. We get 
$< {\cal O} > =  c\, {\imag}(\alpha^*\beta) \, S$, where
$c=0.0050$ at tree level and $c=0.0042$ including the order
$\alpha_s$ corrections. This implies that the effect is too small
in order to obtain an interesting sensitivity to
${\imag}(\alpha^*\beta)$ (i.e.  at the percent level) from
semileptonic top quark decays.

In summary we have computed, for a $tbW$ vertex with left- and
right-handed components, double differential lepton distributions
to order $\alpha_s$ for polarised semileptonic top decay, and we
presented the QCD corrections in terms of compact parameterisations.
These formulae should be a useful module in the theoretical description
of $t$ and/or $\bar t$ production and decay at next-to-leading 
order in $\alpha_s$,
including non-standard top couplings. 

\subsubsection*{Acknowledgments}
We would like to thank A. Brandenburg for useful comments.

\newpage
\begin{center}
FIGURE CAPTIONS
\end{center}
\noindent 
{\bf Fig. 1.} The order $\alpha_s$ contributions to the polarisation 
independent part of the lepton distributions (\ref{xellth}),
(\ref{xnuth}).  \\
{\bf Fig. 2.} The order $\alpha_s$ contributions to the polarisation 
dependent part of the lepton distributions (\ref{xellth}),
(\ref{xnuth}). 

\begin{figure}
\unitlength1.0cm
\begin{center}
\begin{picture}(16,16)
\put(0,0){\psfig{figure=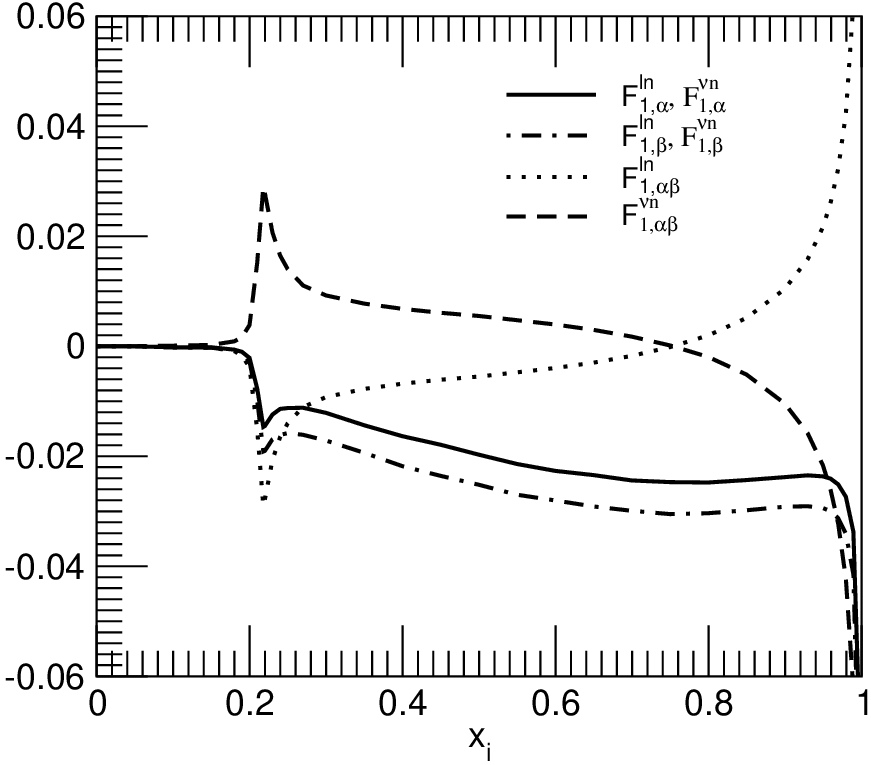,width=12cm,height=12cm}}
\end{picture}
\vskip 0.5cm
\caption{}\label{fig:obs1}
\end{center}
\end{figure}
\begin{figure}
\unitlength1.0cm
\begin{center}
\begin{picture}(16,16)
\put(0,0){\psfig{figure=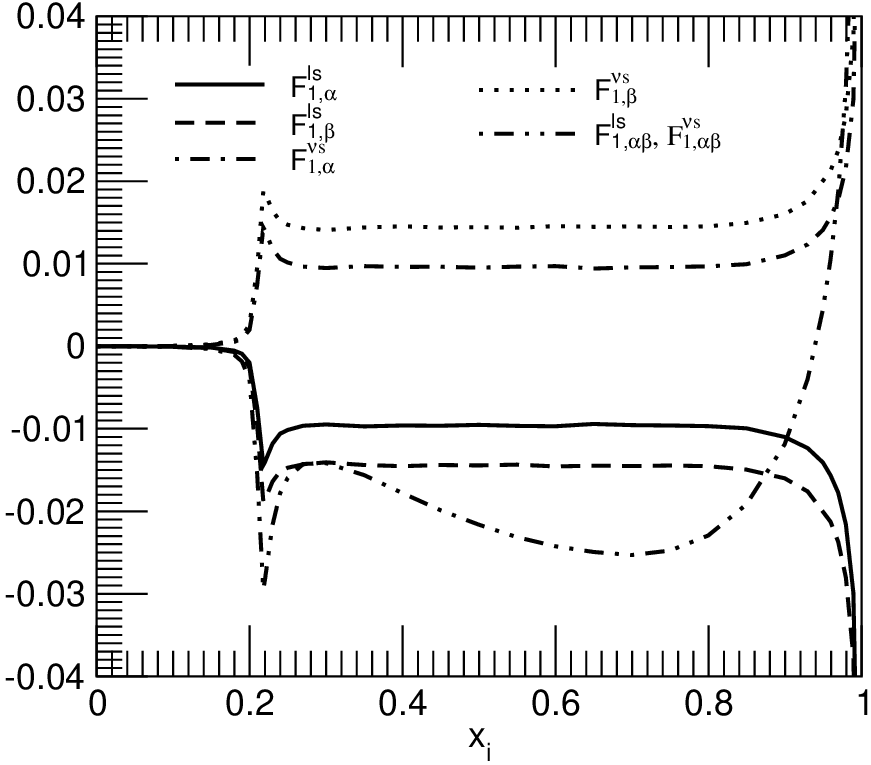,width=12cm,height=12cm}}
\end{picture}
\vskip 0.5cm
\caption{}\label{fig:obs2}
\end{center}
\end{figure}

\end{document}